\documentstyle[preprint,aps]{revtex}
\draft
\begin{document}
\title{Coulomb Zero-Bias Anomaly: A Semiclassical Calculation.}
\author{L. S. Levitov$^{a,b}$ and A. V. Shytov$^b$}
\address{(a) Massachusetts Institute of Technology,
12-112,
77 Massachusetts Ave., Cambridge, MA 02139}
\address{(b) L. D. Landau Institute for Theoretical Physics, 2, Kosygin st.,
Moscow, 117334, Russia}
\maketitle
\begin{abstract}
Effective action is proposed for the problem of Coulomb blocking
of tunneling. The approach is well suited to deal with the
``strong coupling'' situation near zero bias, where perturbation
theory diverges. By a semiclassical treatment, we reduce the
physics to that of electrodynamics in imaginary time, and
express the anomaly through exact conductivity of the system
$\sigma(\omega, q)$ and exact interaction. For the diffusive
anomaly, we compare the result with the perturbation theory of
Altshuler, Aronov, and Lee. For the metal-insulator transition
we derive exact relation of the anomaly and critical exponent of
conductivity.
    \end{abstract}
\pacs{PACS numbers: 72.10.Bg, 73.50.Fq, 73.50.Td\\
{\it to appear in Physical Review Letters}}


\noindent{\bf Introduction.}\hskip4mm
Coulomb correlations in a disordered metal suppress
tunneling  conductivity at low bias, and lead to the  ``zero-bias anomaly''
that   has  been  known
experimentaly since the early sixties\cite{review}.
A perturbation theory of this effect was developed by Altshuler,
Aronov and Lee\cite{AAL}. The theory deals  with  the  diffusive
limit, and shows that the blocking increases at small bias which
leads to a singularity in the tunneling conductivity. The theory
has  been  thoroughly  tested  experimentally  and  found  to be
extremely accurate when the anomaly is a weak feature on the top
of a large constant conductivity.

However, in the past few years there has been interest in the systems with
strong Coulomb effects, such as disordered metals and
semiconductors near  metal-insulator transition\cite{Dynes}. It
was found that the Coulomb anomaly sharply increases near the
transition, and thus it provides a test of
the role of Coulomb correlations in strongly disordered systems.
Another important discovery is the observation by Ashoori et al.
of the Coulomb blocking of tunneling in a two-dimensional metal in
magnetic field\cite{Ashoori}. In this experiment it was found
that at certain magnetic field the zero-bias anomaly abruptly
increases and transforms to a ``soft Coulomb gap''. It has been
pointed out\cite{Eisenstein} that this transition is induced by disorder.
More recently, the gap was studied in the systems with higher
mobility\cite{Eisenstein,gap}, where current is almost entirely blocked
below certain threshold bias. These findings caused a lot
of theoretical work, concerned with the behavior of current near zero
bias\cite{HeEtAl}, and with determining the gap
width\cite{gap}.

It is characteristic for all works cited above that they start
with some model of the system, and then proceed with a many-body
calculation. However, it would be desirable to have an approach
that does not rely on any model, which would allow a comparison
with experiment in the situations where there is no accepted
model for conductivity. The goal of this paper is to propose an
effective action theory that treats the Coulomb anomaly as {\it
cooperative tunneling} and gives tunneling rate in terms of the
actual conductivity $\sigma(\omega,q)$ of the system. For the
first time, it provides non-perturbative treatment of the
anomaly, accurate in the strong coupling regime. For example, at
the metal-insulator transition, conductiviy has scale invariant
form $\sigma(\omega)\sim \omega^\alpha$, and we are able to predict
the form of the tunneling $I-V$ curve.
This relation may be used to  determine the critical  exponent
$\alpha$. We discuss it below, and  also  apply  the  method  to
diffusive  anomaly  in a two dimensional metal, with and without
magnetic  field.  Comparison  with   the   perturbation   theory
results\cite{AAL,HeEtAl} is given, and an agreement is found.

The physics of tunneling involves motion of a large number of
electrons: while only one electron is actually transferred
across the barrier, many other electrons are moving coherently
to accomodate the new electron, and this collective effect
completely controls the tunneling rate. We will argue that this
motion can be treated semiclassically as classical
electrodynamics in imaginary time, find instanton solution, and
derive an expression for the tunneling rate in terms of the
instanton action. To justify the semiclassical picture, let us
consider a situation when at small bias one electron crosses the
barrier. Typically, the time of flight through the barrier is
much shorter than the relaxation time in the electron liquid.
Therefore, while one electron is traversing the barrier other
electrons practically do not move. Thus instantly a large
electrostatic potential is formed, both due to the tunneling
electron itself, and due to the screening hole left behind. The
jump in electrostatic energy by an amount much bigger than the
bias $eV$ means that right after the one electron transfer we
find the system in a classically forbidden state ``under'' the Coulomb
barrier. In order to accomplish tunneling, the charge yet has to
spread over a large area, so that the potential of the charge
fluctuation is reduced below $eV$. If the conductivity is
finite, the spreading over large distance takes long time, and
thus the action of the whole system under-barrier motion is much
bigger than $\hbar$.

An attempt to derive the Coulomb anomaly semiclassically
was made by Spivak\cite{Spivak} who considered a two
dimensional metal, and used Maxwell's theory of charge spreading,
$r(t)\sim\sigma t$, to calculate the Coulomb part of the
action:
  \begin{equation}
{\cal S}(t)\sim \int^t_{t_0}{e^2\over r(t')}dt'=
{e^2\over\sigma}\ln\left({t\over t_0}\right)
\label{Spivak}
\end{equation}
   From that, the tunneling is suppressed by
$\exp\left(-{1\over\hbar}{\cal S}(t_\ast)\right)$, where
$t_\ast=\hbar/eV$. The estimate shows that the action grows at
small bias, which is a clear sign that the semiclassical
treatment is meaningful even for a well conducting metal.
However, in the diffusive limit the estimate (\ref{Spivak}) does
not agree with the perturbation theory. We shall see that the
reason is that the main part of the action is rather Ohmic than
Coulomb, and that after writing the action properly the
semiclassical method completely recovers the pertubation theory
result.

\noindent{\bf The action.}\hskip2mm
  Let us write the action in terms of the charge and current
densities $\rho(r,t)$ and ${\bf j}(r,t)$. The contribution to
the action of the spreading charge is mainly coming from long
times when the charge deviation from equilibrium is small.
Therefore, we can expand the action in powers of $\rho(r,t)$ and
${\bf j}(r,t)$, and keep only quadratic terms. The action should
reproduce the classical electrodynamics equations: the Ohm's law
and charge continuity. In principle, this requirement is
sufficient to determine the form of the action. However, it is
more convenient to argue in the following way. We are going to
use the action to study the dynamics in imaginary time.
Therefore, the action is precisely the one that appears in the
quantum partition function. The latter action expanded up to
quadratic terms in charge and current density must yield
correct Nyquist spectrum of current fluctuations in equilibrium:
   \begin{equation}\label{Nyquist}
\langle\!\langle {\bf g}_{\omega,q}^\alpha
{\bf g}_{-\omega,-q}^\beta \rangle\!\rangle =
\sigma_{\alpha\beta}|\omega|+
\sigma_{\alpha\alpha'}D_{\beta\beta'}q_{\alpha'}q_{\beta'}\ .
   \end{equation}
Here
\begin{equation}
{\bf g}={\bf j}+\hat D\nabla \rho
\end{equation}
  is external current and $D_{\alpha\beta}$ is the tensor of
diffusion constants related to the conductivity tensor by
the Einstein's formula: $\hat\sigma=e^2\nu\hat D$, where
$\nu=dn/d\mu$ is compressibility. Generally,
both $\hat\sigma$ and $\hat D$ are functions of the frequency and momentum.
For simplicity, we assume that
the temperature is zero and discuss only a two dimensional metal
with spatially isotropic and homogeneous conductivity:
$\sigma_{xx}=\sigma_{yy}$, $\sigma_{xy}=-\sigma_{yx}$.

The requirement that the action produces correct current
fluctuations is essentially equivalent to the
fluctuation-dissipation theorem. Thus, in imaginary time we get
   \begin{equation}\label{action}
{\cal S}={1\over 2} \int\!\int\! d^4x_1d^4x_2\!
\left[{\bf g}_{1}^{T}
\hat K_{x_1\!-x_2}{\bf g}_{2}+
{\delta_{12}\rho_{1} \rho_{2}\over|r_1-r_2|}\right]
\end{equation}
where $x_{1,2}=(t_{1,2},r_{1,2})$, and $\delta_{12}=\delta(t_1-t_2)$.
The kernel $\hat K_{r,t}$ is  related  to  the  current correlator,
  \begin{equation}
(K^{-1}_{\omega,q})_{\alpha\beta}=
\langle\!\langle {\bf g}_{i\omega,q}^\alpha
{\bf g}_{-i\omega,-q}^\beta \rangle\!\rangle
\end{equation}
   given by (\ref{Nyquist}), where $\hat\sigma$ and $\hat D$ are
functions of the Matsubara frequency related with
the real frequency functions by the usual analytic continuation.
We take Coulomb interaction in the second term of the
action (\ref{action}) as non-retarded because we are going to study
systems with relatively low conductivity, and thus slow charge relaxation.


\noindent{\bf Dynamics in imaginary time.}\hskip4mm
  To calculate the tunnelling rate, we use the  instanton
method and look for a path in imaginary
time. Among the ``bounce'' paths symmetric in time,
$\rho(r,t)=\rho(r,-t)$, ${\bf j}(r,t)=-{\bf j}(r,-t)$,
we shall find the least action path which will give a
semiclassical estimate of the tunneling rate exponent.

Let us derive equation of motion from the variational
principle. We note that the action (\ref{action}) contains the
charge and current densities as {\it independent} variables, as
Eq.(\ref{action}) was derived by matching with the equilibrium
fluctuations in the {\it grand canonical ensemble} where
charge is not conserved. Therefore, we have to supply the action
(\ref{action}) with the charge continuity constraint:
$\dot\rho+\nabla\cdot{\bf j}={\cal J}(r,t)$, where
\begin{equation}\label{source}
{\cal J}(r,t)=e\delta(r)(\delta(t+\tau)-\delta(t-\tau))\ .
\end{equation}
  This form of the charge source ${\cal J}(r,t)$ describes
electron entering the system at $r=0$,
$t=-\tau$, and exiting at $t=\tau$ through the same point.

By incorporating the charge continuity we get
   \begin{equation}
{\cal S}_{total}={\cal S}(\rho,{\bf j})+\phi(r,t)\left(
\dot\rho+\nabla{\bf j}-{\cal J}(r,t)\right)\ ,
\end{equation}
   where $\phi(r,t)$ is Lagrange multiplier. For the least action path,
the variation
of ${\cal S}_{total}$ relative to infinitesimal change $\delta\phi$,
$\delta\rho$, and $\delta{\bf j}$ vanishes.
(Note that due to the charge
continuity $\delta\dot\rho+\nabla\cdot\delta{\bf j}=0$.)
 After
eliminating $\phi$ we get the   standard equations of classical
electrodynamics:
\begin{eqnarray}\label{dynamics}
({\rm i})\qquad & & \dot\rho+\nabla\cdot{\bf j}={\cal J}(r,t)\ ;\nonumber\\
({\rm ii})\qquad & & {\bf j}+D\nabla\rho =\hat\sigma(\omega,q){\bf E}\ ;\\
({\rm iii})\qquad & & {\bf E}(r,t)=-\nabla_r\int dr'{\rho(r',t)U(|r-r'|)}\ .
\nonumber
\end{eqnarray}
The equations describe the system trajectory in imaginary  time,
i.e.,  the  spreading of charge under the Coulomb barrier due to
selfinteraction.

The next step is  to  solve
Eq.(\ref{dynamics})  for  $\rho$ and ${\bf j}$, and to compute
the action (\ref{action}). For a spatially  homogeneous system,
by using Fourier transform, we get
  \begin{eqnarray*}
\rho(\omega,q)&=&{{\cal J}(\omega)\over |\omega|+Dq^2+\sigma_{xx}q^2U_q}\ ,\\
{\bf j}(\omega,q)&=&{-i\hat K^{-1}(\omega,q){\bf q}U_q \rho (\omega,q)}\ ,
\end{eqnarray*}
where $U_q$ is the Coulomb potential formfactor.
Substituted in Eq.(\ref{action}) this yields the action
\begin{equation}\label{Stau}
{\cal S}_0(\tau)=
{1\over2}\sum\limits_{\omega,q}\,
{|{\cal J}(\omega)|^2\over|\omega|+Dq^2}\,
{U_q\over|\omega|+Dq^2+\sigma_{xx}q^2U_q}
\end{equation}
which  depends  on the accomodation time $\tau$ through Fourier
component of the charge source: ${\cal J}(\omega)=2ie\sin\omega\tau$.

To obtain total action of the system we subtract from the
action ${\cal S}_0(\tau)$ of spreading charge the term $2eV\tau$
that accounts for the work done by voltage source:
\mbox{${\cal S}(\tau) = {\cal S}_0(\tau) -2eV\tau$}.
Thus the energy conservation at
transferring one electron  across the barrier is assured.
Then one has to optimize ${\cal S}(\tau)$
in $\tau$.
Optimal $\tau_\ast$
satisfies the relation
\begin{equation}\label{tau}
\frac{\partial {\cal S}_0(\tau_\ast(V))}{\partial \tau} = 2eV
\end{equation}
  Having solved Eq.(\ref{tau}) for $\tau_\ast$, one obtains
tunneling rate that coinsides with conductivity up
to a constant factor:
  \begin{equation}\label{probability}
G(V)=G_0\exp\left[-{1\over\hbar}\left({\cal S}_0(\tau_\ast(V))
- -2eV\tau_\ast(V)\right)\right]\ .
\end{equation}
The optimal $\tau_\ast $ can be interpreted as the charge accomodation
time.

The accuracy of the term $2eV{\tau}$ is determined by the assumption
that $\tau_\ast\gg\tau_f$, the time it takes one electron to traverse
the barrier. This assumption is valid whenever there is an anomaly:
if $\tau_\ast\approx\tau_f$,
then ${\cal S}(\tau_{\ast})\approx\hbar$, and thus there is no tunneling
suppression.

Taken together, the equations (\ref{Stau}), (\ref{tau}), and
(\ref{probability}) define conductivity. With this general
framework, one can study the anomaly in different systems. Let
us emphasize that after having calculated $\tau_\ast(V)$ and
${\cal S}(\tau_\ast)$ it is essential to check the
selfconsistency of the assumption that $\tau_\ast\gg\tau_f$. To
illustrate this point, let us consider a {\it clean} metal,
where $\sigma(\omega)=ine^2/m\omega$. In this case
Eq.(\ref{Stau}) gives ${\cal S}_0(\tau)\approx\hbar$ at any
$\tau\gg\tau_f$, and henceforth $\tau_\ast\simeq\tau_f$.
Naturally, this indicates absence of the anomaly in a clean
metal, the result familiar from the Fermi liquid picture.

\noindent{\bf Diffusive anomaly.}\hskip 4mm
For a two dimensional metal with elastic scattering time $\tau_0$ and
non-screened Coulomb interaction we set
$U_q = 2\pi/|q|$ and
$\sigma_{xx}=\sigma$,  constant at $|\omega|, v_F|q|\le 1/\tau_0$. Then
Eq.(\ref{Stau}) gives
\begin{equation}\label{AAL}
{\cal S}(\tau) =\,
{e^2\over 8\pi^2\sigma}\,
\ln \left({\tau\over\tau_0}\right)\,
\ln \left(\tau \tau_0 \sigma^2 (\nu e^2)^2 \right)\ .
\end{equation}
(Here $\nu$ is compressibility.) From Eq.(\ref{tau}),
 \begin{equation}\label{AAL_tau}
\tau_\ast = \frac{e}{4\pi^2 V\sigma}\,
\ln(\hbar\sigma\nu e/V)\ .
\end{equation}
The theory is selfconsistent in the hydrodynamic limit,
$\tau_\ast\ge\tau_0$, i.e., at $eV\le e^2/\sigma\tau_0$.
Then the least action is
\begin{equation}
\label{DoubleLog}
{\cal S}(V) = \frac{e^2}{8\pi^2\sigma}\,
\ln\left(\frac{e}{4\pi^2\sigma V\tau_0}\right)\,
\ln\left(\frac{e\tau_0\sigma(\nu e^2)^2}
{4\pi^2 V}\right)
\end{equation}
   It is interesting to compare this result with the identical
double-log dependence derived by Altshuler, Aronov, and Lee in a
different context \cite{AAL}. They calculated perturbatively the
correction to the tunneling density of states
$\delta\nu(\epsilon)$ with the assumption that it is small,
$|\delta\nu|\ll\nu_0$, which is the case only for a
weak disorder. It was found that
$\delta\nu(\epsilon)=-\hbar^{-1}\nu_0{\cal S}(V=\epsilon/e)$,
where ${\cal S}(V)$ is given by (\ref{DoubleLog}). The main
difference is that our double-log has to be exponentiated to get
the tunnelling density of states, while in \cite{AAL} the
double-log itself appears as a correction to the density of
states. In the range of the perturbation theory validity the two
results agree. From that point of view, our calculation provides
description of the diffusive anomaly at low bias, where the
perturbation theory diverges.

\noindent {\bf Screening by electrodes.}
\hskip 4mm
In a real experiment the charge tunnels between two
electrodes, and there are separate contributions to the action due to
the relaxation of the electron and hole charges on both sides of
the barrier. If the electrodes are close, the
charges partially screen the field of each other, which
makes their spreading correlated. In this case the least action
is smaller than the sum of independent contributions of the
electrodes, and thus the anomaly is weakened. For a two
dimensional system, in terms of algebra, the effect will be that
the log-divergence of the integral over $q$ in Eq.(\ref{Stau})
will be cut at $q\simeq a^{-1}$,
where $a$ is the
distance between the electrodes. As a result,
the $V-$dependence of the second log in Eq.(\ref{DoubleLog})
saturates at $eV\simeq V_0=\hbar\sigma/a$.

This ``excitonic'' correlation effect can be treated
straightforwardly by writing the action (\ref{action}) for each
electrode separately, together with the term describing
interaction across the barrier. First, let us consider two
identical parallel planes at distance $a$, and assume $e^2\nu
a\gg1$, which is the case in almost all experiments. Then the
least action is
  \begin{equation}\label{otvet}
{\cal  S}_0(\tau)=\alpha\ln\left({\tau\over\tau_0}\right)\ {\rm at}\
\tau\gg\hbar/ eV_0\ .
\end{equation}
   Here
$\alpha={e^2\over2\pi^2\sigma}\ln2\pi e^2\nu a$.
If the planes have different conductivities and
densities of states, the least action still has the form
(\ref{otvet}), but now
  \begin{equation}\label{alpha}
\alpha={e^2\over4\pi^2}
\left[{1\over\sigma_{1}}\,\ln{4\pi\sigma a\over D_2}
+ {1\over\sigma_{2}}\,\ln{4\pi\sigma a\over D_1}\right]\ ,
\end{equation}
  where
$\sigma=\sigma_{1}\sigma_{2}/(\sigma_{1}+\sigma_{2})$, and the
subscripts 1,2 label the  planes.

We find that at low $V<V_0$ the $I-V$ curve is given by the power
law $I\sim V^{\alpha+1}$. As expected,
the tunneling suppression in this case
is weaker than for the non-screened interaction.

\noindent{\bf Anomaly near metal-insulator transition.}
The effective action theory remains valid even for a highly
disordered system where the Drude model of conductivity does not
work. To illustrate this point, let us consider disordered metal
near metal-insulator transition. The problem has been
extensively studied, and properties of conductivity can be
summarized as
follows\cite{review,BelitzKirkpatrick,Finkelstein}. Static
conductivity of the critical state vanishes, and as function of
frequency it obeys scaling law:
$\sigma(\omega)\sim\omega^{\alpha}$, where the exponent $\alpha$
is constrained by $(d-2)/d\le \alpha<1$. The lower bound
$\alpha=(d-2)/d$ is reached for a non-interacting system, and
interaction shifts $\alpha$ up. It is convenient to write
$\alpha=(d-2)/(d-\zeta)$, $0\le\zeta<2$.

We substitute the scaling form in the action (\ref{Stau}) and
get $S(\tau)\sim \tau^\beta$, where
$\beta=\zeta(d-2)/2(d-\zeta)$. Then the time
$\tau_\ast(V)  \sim V^{-1/(1-\beta)}$, which leads to the tunneling rate
  \begin{equation}\label{Gmit}
G\sim\exp-\left(\bar V/ V \right)^\gamma\ ,\ \
\gamma={\zeta(d-2)\over d(2-\zeta)} \ ,
\end{equation}
with universal $\gamma$. The constant $\bar V$ depends on bare
conductivity, and thus is not universal. Let us emphasize again
that the conductivity of a strongly disordered metal is a
difficult many-body problem which we do not attempt to address
here. The meaning of the result is that it shows how
conductivity, if known, can be used to analyze the anomaly.

\noindent{\bf 2D electron gas in magnetic field.}
It is straightforward to incorporate magnetic field in the
theory by substituting $\sigma_{xx} =
\sigma_{xx}(B)$ in the action (\ref{AAL}). (Note that $\sigma_{xy}$
does not enter.)
As magnetic field increases,
the conductivity drops, and at certain
field it reaches the quantum limit $\sigma_q=e^2/\hbar$.
In this field range the
prefactor $\alpha$ in Eq.(\ref{otvet}) becomes of the order of one, and the
anomaly in the conductivity changes from weak to strong. The
threshold conductivity, according to Eq.(\ref{AAL}), is
  \begin{equation}\label{threshold}
\sigma_c={1\over4\pi^2}{e^2\over\hbar} \ln2\pi e^2\nu a
\end{equation}
  A transition like that was observed by Ashoori et al.\cite{Ashoori} in
the tunneling current from a 3D metal into a 2D electron gas. In
this experiment, the ohmic conductance was measured as function of
temperature, which corresponds to our zero temperature
non-linear current taken at $V\simeq k_B T/e$. The 2D gas was
relatively clean with the zero field mobility corresponding to the elastic
scattering time $\tau_0\simeq 4\cdot 10^{-12}\ s$. The Fermi
energy calculated from the electron density was $E_F\simeq 10\
mV$. By using the result (\ref{threshold}) together with the
Drude-Lorentz model,
$\sigma_{xx}(B)=ne^2/m\tau_0\omega^2_c$ at $\tau_0\omega_c\gg 1$,
one finds that the anomaly hardening transition
corresponds to the cyclotron frequency $\omega^{\ast}_c=(8\pi
E_F/\hbar\tau_0)^{1/2} \simeq 8.0\ mV$. In terms of the field
intensity this is approximately $4.6\ Tesla$ which is quite
close to the transition field reported in Ref\cite{Ashoori}.

It is interesting to note that in a weakly disordered metal with
$E_F\tau_0\gg\hbar$ the threshold field is small:
$\hbar\omega^{\ast}_0\ll E_F$. This means that the transition
occurs well below the field where the Quantum Hall state is
formed. Therefore, our estimate of $\omega^{\ast}_0$ based on
the ``bare'' Drude-Lorentz conductivity is
meaningful and legitimate. On the other hand, to find the
current at very low $V$ one would have to use the conductivity
renormalized by localization and interaction effects.

Finally,  let  us mention a relation with the work
by Halperin, He, and Platzman\cite{HeEtAl} that deals with the
anomaly in the $\nu=1/2$ Quantum Hall state. In this work,  the
problem   was   treated  by  summing  linked  cluster  terms  of
perturbation theory, with the density responce function borrowed
from the Chern-Simons Fermi liquid theory\cite{HalperinLeeRead}.
The anomaly was found to have the form:
  \begin{equation}\label{nu=1/2}
G(V)\sim\exp-V_0/V\ , \ \ V_0=4\pi{e\over\epsilon}\sqrt{\pi n}\ ,
\end{equation}
   where $V\ll V_0$, and $n$ is density. It  is  interesting  to
see how this result can be derived from the effective action. It
has  been  shown\cite{HalperinLeeRead}  that conductivity of the
$\nu=1/2$ state has strong spatial dispersion:  $\sigma_k=A|k|$,
$A=e^2/16\pi\epsilon\sqrt{\pi  n}$.  If this form is inserted in
the action (\ref{Stau}), one  gets  $S(\tau)=\pi\sqrt{2\tau/A}$,
which leads to the tunneling rate (\ref{nu=1/2}).

%
\noindent{\bf Conclusion.}\hskip4mm
We argued that the theory of the Coulomb anomaly in the regime
of strong suppression of tunneling is semiclassical. The
underlying reason is that the transfer of one electron across the
barrier is controlled by cooperative motion of many other
electrons. We treat this motion as classical electrodynamics in
imaginary time, write the action and find instanton trajectory.
Relation with the perturbation theory is discussed and an
agreement is found. The tunneling current is expressed in terms
of the actual conductivity of the system, which is useful in the
situations where there is no accepted model of conductivity.

\acknowledgements
We are grateful to B. L. Altshuler, P. A. Lee, and B. I. Shklovskii for
illuminating discussions and useful suggestions.
Research at the L. D. Landau Institute is supported by the
International Science Foundation grant \#M9M000.


\begin{references}
\bibitem{review} B. L. Altshuler and A. G. Aronov, in:
Electron-Electron Interaction in Disordered Systems,
eds. A. L. Efros and M. Pollak
(North-Holland, 1985); R. C. Dynes and P. A. Lee, Science, vol. 223,
p.355 (1984)
   \bibitem{AAL}
B. L. Altshuler, A. G. Aronov, P. A. Lee, Phys. Rev. Lett.,
{\bf 44}, 1288 (1980)
\bibitem{Dynes} J. M. Valles, R. C. Dynes, and J. P. Garno,
  Phys. Rev. B {\bf 40}, 7590 (1989); Phys. Rev. B {\bf 40},6680 (1989)
\bibitem{Ashoori}
R. C. Ashoori, J. A. Lebens, N.P. Bigelow, R.H. Silsbee,
Phys. Rev. Lett., {\bf 64}, 681 (1990);
Phys. Rev. B {\bf 48}, 4616 (1993)
\bibitem{Eisenstein}
J. P. Eisenstein, L.N. Pfeiffer, K.N. West, Phys. Rev. Lett.,
{\bf 69}, 3804 (1992)
\bibitem{gap} Eisenstein et al., Surf. Sci. {\bf 305}, 393 (1994); preprint
(1994);
Brown et al., Phys.Rev.{\bf B 50}, 15465 (1994)
\bibitem{Spivak} B. Z. Spivak, unpublished (1990)
\bibitem{HeEtAl}
S. He, P. M. Platzman, B. I. Halperin, Phys. Rev. Lett., {\bf 71},
777 (1993)
\bibitem{gap}
S. R. E. Yang and A. H. MacDonald, Phys.Rev.Lett.{\bf 70}, 4110 (1993);
Y. Hatsugai, P. A. Bares, and X.G. Wen, Phys.Rev.Lett.{\bf 71}, 424 (1993);
P. Johansson and J. M. Kinaret, Phys.Rev.Lett.{\bf 71}, 1435 (1993);
A. L. Efros and F. G. Pikus, Phys.Rev.{\bf B 48}, 14694 (1993);
C. M. Varma, A. I. Larkin, and E. Abrahams, Phys.Rev.{\bf B 49}, 13999 (1994);
S. R. Renn and B. W. Roberts, Phys.Rev.{\bf B 50}, 7626 (1994);
I. L. Aleiner, H. U. Baranger, and L. I. Glazman,
Phys.Rev.Lett. (1995)
\bibitem{BelitzKirkpatrick}
D. Belitz, T. R. Kirkpatrick, Rev. Mod. Phys.,{\bf 66}, 261 (1994);
\bibitem{Finkelstein}
A. M. Finkelstein, Sov. Sci. Rev. A. Phys.,{\bf 14}, pp.1-101 (1990)
\bibitem{HalperinLeeRead}
B. I. Halperin, P. A. Lee, N. Read, Phys. Rev. B
{\bf 47}, 7312 (1993)
\end{references}
\end{document}